\documentstyle[preprint,aps]{revtex}
\begin{document}
\draft
\title{ Critical behavior for the dilaton black holes}
\author{ Rong-Gen Cai $^{a, b,}$ \footnote{E-mail: cairg@itp.ac.cn}
and Y. S. Myung $^c$ }
\address{$^a$ CCAST (World Laboratory), P.O. Box 8730, Beijing 
100080, China\\ 
$^b$ Institute of Theoretical Physics, Academia Sinica, P.O. Box 2735, 
Beijing 100080, China \\
$^c$ Department of Physics, Inje University, Kimhae 621-749, Korea}

\maketitle

\begin{abstract}
We study the critical behavior in the black $p$-branes and
four dimensional charged dilaton black holes. 
We calculate the thermodynamic fluctuations in the
various (microcanonical, canonical, and grandcanonical) 
ensembles. It is found that the extremal
limit of some black configurations has a critical point and 
a phase transition takes place from the extremal to nonextremal 
black  configurations.
Some critical exponents are obtained, which  satisfy the scaling
laws. This is related to the fact that the entropy of these black 
configurations is a homogeneous function.

\pacs{PACS numbers: 04.70.Dy, 05.40.+j, 05.70.Fh, 64.60.Fr, 
04.50.+h \\
{\it Keywords}: Phase transition, Critical exponents, Black 
$p$-branes, Dilaton black holes}
\end{abstract}

\section{Introduction}

The Hawking  radiation for a black hole 
is one of the most important achievements of quantum
field theory in curved spacetime [1].
 It resolved the puzzle of the mathematical
analogy between the  black hole mechanics and  thermodynamics [2].
Further it offered the basic concept of
black hole  entropy suggested by Bekenstein [3].
 The establishment of temperature
 and entropy for a black hole  implies that the laws of 
thermodynamics can be applied to understand a black hole [4]. 
However there are some important
 differences between black holes and ordinary thermodynamic systems. For
examples, the Hawking-Bekenstein area formula for 
the black hole entropy is not an extensive quantity;
 some black holes have negative heat
 capacities; the system containing a black hole has only a local
equilibrium temperature, which diverges at the black hole horizon.
Unlike the ordinary thermodynamic system, black hole is not a uniform
 system. This means that a black hole cannot be divided into some 
 subsystems. These differences give rise to some difficulties in 
studying the black hole thermodynamics. 
Hence it is of interest to investigate the phase transition in
 black hole thermodynamics.

Nowadays there exist two approaches to the phase
transition of black holes. One  comes form the work of Davies [5].
He argued that a second-order phase transition can take place at the
 singular points of heat capacity of black holes. Many authors have 
 discussed the
 singular points for various black holes [6,7,8,9,10].   Lousto [9]
claimed that the singular points  satisfy the scaling laws of critical
points. In particular,  an effective spatial
dimension of black holes  is obtained as two. Further he insisted 
that his result agrees with the membrane picture of black holes [11]. 
 Later on, it turns out 
that his calculation  is wrong and the
effective spatial dimension of black holes is not two. 
It seems that all
 discussions on his effective two-dimensional quantum
 description of black holes become invalid.

On the other hand, the other group considered that a phase transition
 of black holes occurs when a nonextremal black hole 
 approaches its extremal counterpart. This
 was first reported by Curir [12].   Pav\'on and Rub\'\i\ [13]
found that all second moments of nonequilibrium fluctuations are finite
 for the Schwarzschild black hole with single horizon, but some second 
moments diverge for the extremal
Kerr black hole and extremal Reissner-Nordstr\"{o}m black hole. The
 divergence of second moments means  that a second-order
 phase transition takes place when the
nonextremal black hole is turned into the  extremal one.
 An important point to remark is that all second moments
have no any special behavior and are finite at the points of Davies. 
Cai, Su, and Yu [14]
 obtained  the same result  in the charged dilaton black holes and
 Kerr-Newman black holes. However, all second moments  are finite for
 the $a=1$ charged dilaton black holes.
 So the difference of outer horizon and inner horizon 
plays an important role in the divergence of second
moments, because the $a=1$ charged dilaton black hole, as the 
Schwarzschild black hole, has a single horizon.

Davies' phase transition (that is, due to the
discontinuity of heat capacity) is not so attractive.
This is because the event horizon does not lose
 its regularity and the internal states of black holes are not 
 changed significantly
 through the Davies' points [7].
 On the other hand, it is more interesting to study the phase 
 transition between the extremal and nonextremal black holes.  
 As is well known, the extremal black holes have vanishing Hawking 
 temperature. Therefore, the Hawking evaporation is terminated  
 and only the superradiation is left for an
 extremal black hole. The geometric structure of extremal black holes is
also different with  nonextremal black holes, since the singularity
will be naked beyond the extremal limit.
Wilczek and his collaborators [15] argued that the thermodynamic 
description is inadequate for some extremal black holes and their 
behavior resembles  the normal elementary particles, strings, or
extended objects.  In addition,
 Kaburaki and his collaborators [16] found that the Davies' points 
 in fact
 are turning points, which are related to the changes of stability. 
 Also they claimed that
the divergence of heat capacity at those points does not mean the
occurrance of phase transition.
 More recently, Kaburaki [17] discussed the equilibrium thermodynamic
 fluctuations of Kerr-Newman black hole in various ensembles. He found
that the extremal Kerr-Newman black hole has a critical point and some
critical exponents also obey the scaling laws [18].

 In the previous work [19], by investigating the critical behavior
 and phase transition in the 2+1 dimensional black holes
  recently found by Ba\~nados, 
Teitelboim, and Zanelli (BTZ) [20], we obtained  some interesting
results: (1) The extremal spinning BTZ black hole  has a critical
point
and a second-order phrase transition  takes place from extremal to
nonextremal one. (2) The massless BTZ black hole has
also a critical point and a second-order phase transition occurs from a 
 massless  to massive one. (3) Some critical exponents remain
unchanged (as those of Kerr-Newman black hole) and satisfy the scaling
laws. (4) In particular, the effective spatial dimension of BTZ black
hole is found to be one. This supports   that
extremal black holes are the Bogomol'nyi saturated string states.

On the basis of above results, some questions  arise as follows. Are 
these critical exponents of the extremal black holes universal? Is the
effective spatial dimension of other black holes also one? Could we
obtain a reduced quantum model of black holes? In the present paper, we 
try to reply in
part to the first two questions by extending our discussion to the
black $p$-branes and four-dimensional 
dilaton black holes with (and without) moduli field.
 We note that the thermodynamic properties of these black configurations
  are significantly affected by the dilaton  and moduli fields.

The organization of this paper is as follows. We first
introduce
 the black $p$-branes,  and then discuss their thermodynamic
 properties in Sec. II.
In Sec. III we investigate the equilibrium  fluctuations
of black $p$-branes in the microcanonical ensemble, canonical
 ensemble, and grand canonical ensemble. Further we calculate 
some relevant critical exponents. We discuss the case of the
  four-dimensional charged
 dilaton black holes in Sec. IV. In Sec. V we study the effects of
   a moduli field on the critical behavior of charged dilaton black 
   holes. Finally  we conclude in Sec. VI with a brief discussion.

\section{black strings and $p$-branes  and their thermodynamic
properties}

\subsection{Black strings and Black  $p$-branes}

The $p=1$ black string  and
$p$-brane solutions are relevant to black hole physics [21].
These are some extended objects surrounded by event
horizons.
In this paper, we have $p=\tilde d -1$ for a ``magnetically'' charged 
object and $p= d -1$ for a ``electrically'' charged one.
As black holes in general relativity, these extended objects in the 
string theories  have thermodynamic  properties, such as the 
entropy and Hawking temperature, etc.
 It is  expected that these extended objects  might play the
 important roles in understanding the fundamental nature of sting 
 theory and quantum gravity.
It was shown that they are  solitonic solutions [22], and are closely 
related to fundamental strings 
and membranes themselves [21].
More recently, it appears that the Bekenstein-Hawking
entropy of extremal, near-extremal black holes, and black $p$-branes
agrees with the string entropy derived from the bound states of
 Dirichlet branes [23].
Therefore,  it is
important to discuss the critical behavior for these extremal
black extended objects.

We start with the R-R sector of supergravity action [21,22,24]
\begin{equation}
I_{\rm D}=\frac{1}{16\pi}\int d^Dx\sqrt{-g}\left( R-
\frac{1}{2}(\nabla \phi)^2 -\frac{e^{-\alpha (d)\phi}}{2(d+1)!}
F^2_{d+1}\right)-\frac{1}{8\pi}\int d^{D-1}x\sqrt{-h}K,
\end{equation}
where
$$\alpha ^2(d)=4-\frac{2d\tilde{d}}{d+\tilde{d}},\ \
 \tilde{d}=D-d-2. $$
Here $R$ is the scalar curvature, $\phi$ is the dilaton field, 
$F_{d+1}$ denotes
 the ($d+1$)-form  tensor field and $K$ the extrinsic curvature on
the induced metric $h$. Varying the action (1), one finds the 
equations of motion
\begin{eqnarray}
& &R_{\mu\nu}-\frac{1}{2}g_{\mu\nu}R=\frac{1}{2}\nabla _{\mu}\phi
\nabla _{\nu}
\phi +\frac{1}{2d!}e^{-\alpha \phi}F_{\mu ...}F_{\nu}^{\ ...}
\nonumber\\
    & &-\frac{1}{4}g_{\mu\nu}\left((\nabla  \phi)^2
    +\frac{1}{(d+1)!}e^{
    -\alpha \phi}F^2_{d+1}\right),\\
& &\nabla ^2\phi=\frac{-\alpha}{2(d+1)!}e^{-\alpha \phi}
F^2_{d+1},\\
& &0=\nabla ^{\mu}(e^{-\alpha \phi}F_{\mu ... \nu}).
\end{eqnarray}
 We wish to obtain the general $ (\tilde{d}-1)$-brane solution for
  the action (1).
 First we require that the solution is invariant under the symmetry 
 group $R\times SO(d+2)\times E(\tilde{d}-1)$,
 where $E(n)$ denotes the $n$-dimensional Euclidean group. This is 
 realized as 
\begin{equation}
ds^2=e^{2X}d\hat{s}^2+e^{2Y}dx^idx^i,
\end{equation}
where $d\hat{s}^2$ is a $(d+3)$-dimensional Lorentzian line element,
 $x^i$ are  external coordinates for ($\tilde{d}-1$)-dimensions.  
 All fields and
 functions $X$ and $Y$ are independent of $x^i$ and therefore the 
 solution is invariant under the translational and
rotational transformations in ($\tilde{d}-1)$-dimensions.  Substituting 
 (5) into (1), we obtain the action per unit volume for the 
 ($\tilde{d}-1$)-brane
\begin{equation}
I_{d+3}=\frac{1}{16\pi}\int d^{d+3}x\sqrt{-g}\left (R-\frac{1}{2}
(\nabla\rho)^2-\frac{1}{2}(\nabla \sigma)^2
-\frac{1}{2(d+1)!}e^{\beta \rho}F_{d+1}^2\right ),
\end{equation}
where the $\rho$ and $\sigma$ are defined by the relation
\begin{eqnarray}
\beta X&=&-\frac{d(\tilde{d}-1)}{(d+\tilde{d})(d+1)}\rho +\sqrt{
\frac{\tilde{d}-1}{2(d+\tilde{d})(d+1)}}\alpha (d)\sigma,
\nonumber\\
\beta Y&=&\frac{d}{d+\tilde{d}}\rho-
\sqrt{\frac{d+1}{2(d+\tilde{d})(\tilde{d}+1)}}\alpha (d)\sigma,
 \nonumber\\
\beta \phi&=&-\alpha (d)\rho
-\frac{2d}{\sqrt{2(d+\tilde{d})}}\frac{\tilde{d} -1}{d+1}\sigma,
\nonumber
\end{eqnarray}
and the constant $\beta$ is given by
$$\beta=-\sqrt{\frac{2(d+2)}{d+1}}.$$
The problem of finding black $(\tilde d -1)$-brane solution of (1) is
 thereby reduced to 
the problem of finding black hole solution of (6).
The asymptotically flat and spherically symmetric  black
 hole
 solutions for the action (6)  were found  in [25] as
\begin{eqnarray}
F&=&Q_m\epsilon _{d+1},\\
d\hat{s}^2&=&-\left [1-\left (\frac{r_+}{r}\right)^d\right]\left[1-
\left (\frac{r_-}{r}\right)^d\right]^{1-\gamma d}dt^2
+\left[1-\left (\frac{r_+}{r}\right )^d\right]^{-1}\left[1-
\left(\frac{r_-}{r}\right)^d\right]^{\gamma-1}dr^2
\nonumber\\
&+&r^2\left[1-\left(\frac{r_-}{r}\right)^d\right]^{\gamma }
d\Omega^2_{d+1},\\
e^{\beta \rho}&=&\left[1-\left(\frac{r_-}{r}\right)^d\right]^
{\gamma d},\\
\sigma&=&0,
\end{eqnarray}
where  $\epsilon _{d+1}$ is the volume element of the unit 
$(d+1)$-sphere and the constant $\gamma$ is
$$\gamma=\frac{d+2}{d(d+1)}. $$
Here the constant $Q_m$ is the magnetic charge of the black hole 
defined by two parameters  $r_-$ and $r_+ (\geq r_-)$,
$$ Q_m=d(r_-r_+)^{d/2}.$$
The structure of solutions (7)-(10) is clear.
The parameter $r_+$ corresponds to the event
horizon of the black hole, while  $r_-$ is a scalar curvature 
singularity. It is straightfoward to show that the curvature $R$ 
 is finite at $r=r_+$. 
From the last term in (8) with $\gamma > 0$,
 at $r=r_-$ the area of the $(d+1)$-sphere goes to zero and thus 
$R$ is singular.
In fact  $r=r_-$ is a spacelike surface of singularity,
 very similar to the singularity in the Schwarzschild metric.
Using (5)-(10), we obtain   the
black ($\tilde{d}-1$)-brane solutions for the action (1)
\begin{eqnarray}
&&F=Q_m\epsilon _{d+1},\\
&&e^{-2\phi}=\left[1-\left (\frac{r_-}{r}\right)^d\right]^{
\gamma _{\phi}},\\
&&ds^2=-A(r)dt^2+B(r)dr^2+r^2C(r)d\Omega ^2_{d+1}+D(r)dx^idx^i,
\end{eqnarray}
where
\begin{eqnarray}
A(r)&=&\left [1-\left(\frac{r_+}{r}\right )^d\right ]\left [
1-\left (\frac{r_-}{r}\right )^d\right ]^{\gamma _x -1}, \nonumber\\
B(r)&=&\left[1-\left(\frac{r_+}{r}\right)^d\right]^{-1}\left[1-\left(
\frac{r_-}{r}\right)^d\right]^{\gamma _{\Omega}-1}, \nonumber\\
C(r)&=&\left [1-\left (\frac{r_-}{r}\right )^d\right ]^
{\gamma _{\Omega }},\nonumber\\
D(r)&=&\left [1-\left (\frac{r_-}{r}\right )^d\right ]^{\gamma _x},
\nonumber
\end{eqnarray}
and
$$\gamma _x=\frac{d}{d+\tilde{d}},\ \
\gamma _{\Omega}=\frac{\alpha ^2(d)}{2d},\ \
\gamma _{\phi}=\alpha (d). $$
This is a ``magnetically'' charged black $(\tilde{d}-1$)-brane  solution
 with two parameters $r_-$ and $r_+$. To obtain the black $p$-brane
solutions with ``electric'' charge, we dualize these forms as in [21].
The equation of motion of $F_{d+1}$ in (4)  implies that
the ($\tilde{d}+1$)-form
$\tilde{F}_{\tilde{d}+1}=e^{-\alpha \phi}*F_{d+1}$ is closed, where $*$
 denotes the Hodge dual with $(*)^2=1$. If one replaces $F^2_{d+1}$
by $\tilde{F}_{{\tilde{d}+1}}^2$, $-\alpha $ by $\alpha $, and $(d+1)$
by $(\tilde{d}+1)$, the action (1) then becomes
\begin{equation}
\tilde{I}_{\rm D}=\frac{1}{16\pi}\int d^Dx\sqrt{-g}\left(R-\frac{1}{2}
(\nabla \phi)^2-\frac{e^{\alpha \phi}}{2(\tilde{d}+1)!}\tilde{F}_{
\tilde{d}+1}^2\right)
-\frac{1}{8\pi}\int d^{D-1}x\sqrt{-h}K.
\end{equation}
Considering the action (14), we obtain the equations of motion
\begin{eqnarray}
R_{\mu\nu}&-&\frac{1}{2}g_{\mu\nu}R=\frac{1}{2}\nabla _{\mu}\phi
\nabla _{\nu}
\phi +\frac{1}{2\tilde{d}!}e^{\alpha \phi}\tilde{F}_{\mu ...}
\tilde{F}_{\nu}^{...} \nonumber\\
&-&\frac{1}{4}g_{\mu\nu}\left((\nabla \phi)^2+\frac{1}{(\tilde{d}+1)!}
e^{\alpha \phi}\tilde{F}^2\right),\\
\nabla ^2\phi &=&\frac{\alpha}{2(\tilde{d}+1)!}e^{\alpha \phi}
\tilde{F}^2,\\
0&=&\nabla ^{\mu}(e^{\alpha \phi}\tilde{F}_{\mu ...\nu}).
\end{eqnarray}
Comparing Eqs.(15)-(17) with Eqs. (2)-(4), it is easy to find that Eqs.
 (15)-(17)
are exactly identical with Eqs. (2)-(4), after  replacing 
 $\tilde{F}^2_{\tilde{d}+1}$ by $F^2_{d+1}$, $\alpha$ by $-\alpha$,
and $\tilde{d}$ by $d$. Thus one can  obtain the ``electrically''
 charged black $(d-1)$-brane solutions for the action (1),
\begin{eqnarray}
&&\tilde{F}_{\tilde{d}+1}=\tilde{d}(r_+r_-)^{\tilde{d}/2}\epsilon _{
\tilde{d}+1}\\
&&e^{-2\phi}=\left[1-\left (\frac{r_-}{r}\right)^{\tilde{d}}\right]^{
\tilde{\gamma} _{\phi}},\\
&&ds^2=-\tilde{A}(r)dt^2+\tilde{B}(r)dr^2
+r^2\tilde{C}(r)d\Omega ^2_{\tilde{d}+1}+\tilde{D}(r)dx^jdx^j,
\end{eqnarray}
where
\begin{eqnarray}
\tilde{A}(r)&=&\left [1-\left(\frac{r_+}{r}\right )^{\tilde{d}}\right
 ]\left [
1-\left (\frac{r_-}{r}\right )^{\tilde{d}}\right ]^{\tilde{\gamma} _x
 -1},
\nonumber\\
\tilde{B}(r)&=&\left[1-\left(\frac{r_+}{r}\right)^{\tilde{d}}
\right]^{-1}\left[
1-\left(\frac{r_-}{r}\right)^{\tilde{d}}\right]^{\tilde{
\gamma} _{\Omega}-1},
\nonumber\\
\tilde{C}(r)&=&\left [1-\left (\frac{r_-}{r}\right )^{\tilde{d}}
\right ]^{\tilde{\gamma} _{\Omega }},\nonumber\\
\tilde{D}(r)&=&\left [1-\left (\frac{r_-}{r}\right )^{\tilde{d}}
\right ]^{\tilde{\gamma} _x}, \nonumber
\end{eqnarray}
and
$$\tilde{\gamma} _x=\frac{\tilde{d}}{d+\tilde{d}},\ \
\tilde{\gamma} _{\Omega}=\frac{\alpha ^2(d)}{2\tilde{d}},\ \
\tilde{\gamma} _{\phi}=-\alpha (d). $$
The ``electric'' charge $Q$ per unit volume of the black
 $(d-1)$-branes can be expressed in terms of $r_-$ and
  $r_+(\geq r_-) $
\begin{equation}
Q=\frac{\Omega _{\tilde{d}+1}}{4\pi}\tilde{d}(r_+r_-)^{\tilde{d}/2},
\end{equation}
where $\Omega _{\tilde{d}+1}$ is the volume of the unit
 ($\tilde{d}+1)$-sphere.
 One of the main goals of this work is to investigate the
critical behavior in the  black $p$-brane solutions with
 ``electric'' charge.

\subsection{Thermodynamics}

In order to investigate the critical behavior of black strings
 and $p$-branes,
 some relevant thermodynamic quantities must be derived first. The
 thermodynamics of these black configurations has been discussed by us
 and
Muto [26], but  a few misprints and calculation mistakes exist in [27].
 For
the completeness,  here we list some main results of thermodynamics of
 these objects. Analytically continuing the solution (20) to
 its Euclidean sector leads to
\begin{equation}
ds^2_{\rm E}=\tilde{A}(r)d\tau ^2+\tilde{B}(r)dr^2+r^2\tilde{C}(r)
d\Omega^2_{\tilde{d}+1}
+\tilde{D}(r)dx^jdx^j,
\end{equation}
where $\tau$ is the Euclidean time.  We can  obtain the Hawking 
temperature $T$ by requiring the absence of the conical
 singularity in the Euclidean spacetime
\begin{equation}
T \equiv \beta _H^{-1}=\left.\frac{\tilde{A}'}{4\pi\sqrt{\tilde{A}
\tilde{B}}}\right |_{r_+}\nonumber\\
=\frac{\tilde{d}}{4\pi r_+}\left[1-\left(\frac{r_-}{r_+}
\right)^{\tilde{d}}\right]^{(\tilde{d}-2)/2\tilde{d}},
\end{equation}
where a prime represents derivative with respect to $r$. The 
Hawking temperature
(23) can also be derived from the surface gravity $\kappa$ 
($T=\kappa /2\pi$) at the event
horizon
$r_+$, which satisfies
\begin{equation}
l^{\mu}_{;\nu}l^{\nu}=\kappa l^{\mu}.
\end{equation}
Here $l^{\mu}$ is a timelike Killing vector of solution (20).
 From (14) the Euclidean action
of black ($d-1$)-brane is
\begin{equation}
\tilde{I}_{\rm DE}=-\frac{1}{16\pi}\int d^Dx\sqrt{g}
\left(R-\frac{1}{2}(\nabla \phi)^2-\frac{e^{\alpha \phi}}
{2(\tilde{d}+1)!}\tilde{F}^2_{\tilde{d}+1} \right)
+\frac{1}{8\pi}\int d^{D-1}x\sqrt{h}(K-K_o),
\end{equation}
where $K_o$ is the extrinsic curvature of a fixed $r > r_+$ timelike
supersurface
embedded into the flat D-dimensional Minkovski spacetime. We have to
 introduce this term in order to normalize the Euclidean action to 
 zero for a flat Minkovski spacetime [28]. For the Euclidean 
 spacetime (22), we have
\begin{eqnarray}
R=-\frac{1}{\sqrt{g}}\left (\frac{\sqrt{g}\tilde{A}'}
{\tilde{A}\tilde{B}}\right)'-2G_0^{\ 0}, \nonumber \\
K=-\frac{1}{\sqrt{g}}\left( \frac{\sqrt{g}}{\sqrt{
\tilde{B}}}\right)', \ \ K_o=-\frac{\tilde{d}+1}{r},
\end{eqnarray}
where $G_0^{\ 0}$ is the $00$-component of the Einstein tensor.
Substituting (26) into (25) leads to the Euclidean action per
unit ($d-1$)-brane volume
\begin{equation}
\tilde{I}_{\rm DE}=\frac{\Omega _{\tilde{d}+1}}{16\pi}[(\tilde{d}+1)
r_+^{\tilde{d}}-r_-^{\tilde{d}}]
\beta _H -\frac{1}{4}\Sigma-\frac{\pi Q}{\tilde{d}\Omega _{\tilde{d}+1}
r_+^{\tilde{d}}}
Q\beta _H
\end{equation}
where $\Sigma$ is the horizon area of the black ($d-1$)-brane,
\begin{equation}
\Sigma=\Omega _{\tilde{d}+1}r_+^{\tilde{d}+1}\left [1-\left(\frac{r_-}
{r_+}\right)^{\tilde{d}}\right] ^{(\tilde{d}+2)
/2\tilde{d}}.
\end{equation}
Comparing the action (27) with the thermodynamic potential of an 
ordinary thermodynamic system [29], we obtain the 
Arnowitt-Deser-Misner (ADM) mass
 $M$, entropy $S$, and the chemical potential $\Phi$ with 
 charge $Q$,
\begin{eqnarray}
M&=&\frac{\Omega _{\tilde{d}+1}}{16\pi}[(\tilde{d}+1)r_+^{\tilde{d}}
-r_-^{\tilde{d}}],\\
S&=&\frac{1}{4}\Sigma=\frac{1}{4}\Omega _{\tilde{d}+1}r_+^{\tilde{d}
+1}\left [1-\left(\frac{r_-}{r_+}\right)^{\tilde{d}}\right] ^{
(\tilde{d}+2)/2\tilde{d}},\\
\Phi &=&\frac{\pi}{\tilde{d}\Omega _{\tilde{d}+1}}\frac{Q}{r_+^{
\tilde{d}}}=\frac{1}{4}\left(\frac{r_-}
{r_+}\right)^{\tilde{d}/2}.
\end{eqnarray}
We note that the chemical potential $\Phi$ is just the electric 
potential at the
horizon $r_+$. The ADM mass formula (29) was also obtained in [24] 
by using the gravitational energy-momentum pseudotensor.
Varying the ADM mass, one can obtain the first law of thermodynamics 
for the black extended objects
\begin{equation}
dM=\beta _H^{-1}dS+\Phi dQ,
\end{equation}
and its integration form is
\begin{equation}
M=\frac{\tilde{d}+1}{\tilde{d}}\frac{\kappa}{2\pi}\frac{\Sigma}{4}+
\Phi Q.
\end{equation}
According to the formula $C_Q=(\partial M/\partial T)_Q$, the  heat
capacity is calculated as
\begin{equation}
C_Q=\frac{r_+ \Omega _{\tilde{d}+1}
[(\tilde{d}+1)r_+^{\tilde{d}}+
r_-^{\tilde{d}}][1-(r_-/r_+)^{\tilde{d}}]^{(2+\tilde{d})/2\tilde{d}}}
{4[(\tilde{d}-1)r_-^{\tilde{d}}/r_+^{\tilde{d}}-1]} .
\end{equation}

Now let us discuss the thermodynamic properties of these black 
configurations. At the extremal limit ($r_-=r_+$)  the Hawking 
temperature is zero, finite,
and divergent  when $\tilde{d} > 2$, $\tilde{d}=2$, and 
$\tilde{d} < 2$ (i.e., $\tilde{d}=1$), respectively. But the entropy 
 always vanishes. The heat
capacity (34) approaches zero in the extremal limit for a general
$\tilde{d}$, whereas it is finite when $\tilde{d}=2$. We notice that
 the heat capacity is always negative for  $\tilde{d}=1$.  When
\begin{equation}
(\tilde{d}-1)r_-^{\tilde{d}}-r_+^{\tilde{d}}=0,
\end{equation}
the heat capacity diverges.
 If $(\tilde{d}-1)r_-^{\tilde{d}} > r_+^{\tilde{d}}$, the heat
 capacity is positive and  otherwise it is negative. Therefore, 
 the points satisfying
 equation (35) correspond to the critical points of Davies. 
 The point of
Davies does not exist in the $p$-branes when $\tilde{d}=1$ and
$\tilde{d}=2$. In addition, it is worth noting that
the ADM mass $M$, ``electric'' charge $Q$, entropy $S$, and heat
 capacity $C_Q$
are the extensive quantities with respect to the volume of 
$(d-1)$-branes.
 Thanks to this, the  entropy of near-extremal non-dilatonic
 $p$-branes may
 be described by free massless fields on the world volume [23]. It
 should be pointed out that
 these  quantities only involve the dimensionality ``$\tilde{d}$ '', 
 but not the dimensionality ``$d$ ''. Finally, it is noted
 that the thermodynamics of the ``magnetically'' charged
 black $(\tilde{d}-1)$-branes can
be obtained from the  ``electrically'' charged black $(d-1)$-branes
 by the transformation $\tilde{d}\rightarrow d$. Hence their 
 properties are  similar to those of the ``electrically'' charged 
black  $p$-branes.

\section{Equilibrium thermodynamic fluctuations and 
critical behavior of black $p$-branes}

\subsection{Fluctuation theory of equilibrium 
thermodynamics}

In order to discuss the equilibrium thermodynamic fluctuations
 of black strings and $p$-branes, we first  briefly review 
 equilibrium fluctuation theory in specified environments.
  This was employed to study the critical behavior of
 Kerr-Newman black holes and BTZ black holes. In a 
 self-gravitating thermodynamic system, the thermodynamic 
 fluctuations are different in accordance with the different
environments. Also thermodynamic stability depends on the 
chosen environment [30].
 Hence, we must study the
thermodynamic fluctuations of black configurations in 
 various ensembles.

In general, the equilibrium state of a thermodynamic system can be
 described
completely by the Massieu function  $\Psi$ (the characteristic
thermodynamic function). 
Let us consider
 the infinitesimal variation of the characteristic function [17],
\begin{equation}
d\Psi =\sum ^{n}_{i=1}X_idx_i,
\end{equation}
where the $n$ is the number of the variables $x_i$.
 $\{x_i\} $ is  a set of the intrinsic variables $x_i$ and are
  specified directly
by the environments of  thermodynamic system. 
  $\{X_i\}$ is a set of
variables conjugate to the set $\{x_i\}$, and can be expressed as  
functions of their intrinsic variables $\{x_i\}$:
$$ X_i=\left(\frac{\partial \Psi}{\partial x_i}
\right)_{\bar{x}_i}.$$
Here the subscript $\bar{x}_i$ denotes a set of variables 
$(x_1,\cdots,x_{i-1},x_{i+1},\cdots, x_n)$ exclusive of
 $x_i$. Then the deviation from the equilibrium
state can be described by the  distribution function
\begin{equation}
P(\xi _1,\cdots,\xi _n)d\xi_1 \cdots d\xi_n \sim 
\exp[k_B^{-1}(\hat{\Psi}-\Psi)]
d\xi_1 \cdots d\xi_n,
\end{equation}
where $k_B$ is the Boltzmann's constant.
 $\xi_i\equiv \delta X_i$ stands
 for the deviation of the $i$th conjugate variable from its 
 equilibrium value. And $\hat{\Psi}$ represents the Massieu
  function which is analytically continued to the
 nonequilibrium points near the equilibrium sequence in the
  phase space 
($\{x_i\}, \{X_i\}$). Considering this up to  second order, 
one has [17]
\begin{equation}
P(\{\xi_i\})\sim \exp[-(2k_B)^{-1}\sum^{n}_{i=1}\lambda _i \xi ^2_i],
\end{equation}
where $\lambda _i=(\partial x_i/\partial X_i)_{\bar{x}_i}=
(\partial^2 \Psi /\partial x_i^2)^{-1}_{\bar{x}_i}$ are the eigenvalues
 of the fluctuation modes $\delta X_i$. It follows from (38) that
 the average of each fluctuation mode is zero, but the second moments
  are given by
\begin{equation}
\langle\delta X_i\delta X_j\rangle=\frac{k_B}{\lambda _i}\delta _{ij}
=k_B\left(\frac{\partial ^2\Psi}{\partial x^2_i}\right)_{\bar{x}_i}
\delta _{ij}.
\end{equation}
From (39) we can see that the second moments diverge only
if $\lambda _i=0$,
i.e., $(\partial X_i/\partial x_i)_{\bar{x}_i}=\pm \infty$.
 According to the turning point method of stability 
analysis developed by Katz [31], the change
 of stability occurs only at the turning points where the tangent
 of the curve changes its sign through an infinity (a vertical 
tangent) in the ``conjugate 
diagrams'', in which the conjugate variables $\{X_i\}$ are plotted 
against an intrinsic variable (other intrinsic variables are fixed).
The Davies' points
 are just such turning points. The divergent behavior of some
  second moments at the
Davies' points  means only that the stability for Kerr-Newman
black holes is changed
 [16]. However this has nothing to do with the second-order phase
 transitions.
 On the other hand the divergence of some second moments at the 
 extremal point is related to
the second-order phase transition.
This means that the extremal point is different from the turning
point [17,19]. Therefore in the rest of paper we will  explore
 the behavior of black configurations near the extremal limit.

\subsection{Equilibrium fluctuations of black strings and
 $p$-branes}

Here we investigate
 the equilibrium thermodynamic fluctuations of black strings and
$p$-branes in the various ensembles.

(i) {\it Microcanonical ensemble}. In this ensemble,
 the proper Massieu function of a thermodynamic system is just 
 the entropy of the system. For the black
 strings and $p$-branes,  their entropy is given by the
equation (30) and its  variation can be obtained from  (32)
\begin{equation}
d\Psi _1=dS=\beta _HdM-\beta _H\Phi dQ.
\end{equation}
Comparing (40) with (36), we have two intrinsic variables 
$\{x_i\}=\{M, Q\}$
 and the conjugate variables $\{X_i\}=\{\beta _H, -\varphi\}$ with
  $\varphi=\beta _H\Phi$.
 The eigenvalues for 
 $\beta _H$ and $\varphi$ are given by
\begin{eqnarray}
\lambda_{1m}&=&\left(\frac{\partial M}{\partial \beta _H}\right)_Q=
-T^2C_Q,\\
\lambda _{1q}&=&-\left(\frac{\partial Q}{\partial \varphi}\right)_M=
-TK_M,
\end{eqnarray}
where $C_Q$ is the heat capacity given by (34) and
\begin{eqnarray}
K_M &\equiv &\beta _H\left(\frac{\partial Q}{\partial 
\varphi}\right)_M
=\frac{\tilde{d}\Omega _{\tilde{d}+1}}{\pi}\left (\frac{r_+}{r_-}
\right)^{\tilde{d}}[(\tilde{d}+1)r_+^{\tilde{d}}+
r_-^{\tilde{d}}] \nonumber\\
&\times&\{[(\tilde{d}+1)r_+^{\tilde{d}}-r_-^{\tilde{d}}]+
[(\tilde{d}-1)-(r_-/r_+)^{\tilde{d}}]
[1-(r_-/r_+)^{\tilde{d}}]^{-1}\}^{-1},
\end{eqnarray}
is the ``electric capacitance'' of the black $p$-branes. It
follows from (39) that
\begin{eqnarray}
\langle\delta \beta _H\delta \beta _H\rangle&=&-k_B
\frac{\beta ^2_H}{C_Q},\nonumber\\
\langle\delta \varphi \delta \varphi\rangle&=&-k_B
\frac{\beta _H}{K_M},\nonumber\\
\langle\delta \Phi\delta \Phi\rangle&=&-k_B
\left(\frac{T}{K_M}+\frac{\Phi^2}{C_Q}\right),
\nonumber\\
\langle\delta \beta _H \delta \Phi\rangle&=& k_B
\frac{\beta _H \Phi}{C_Q}.
\end{eqnarray}
  $\lambda_{1m}$ and
 $\lambda _{1q}$ in (41) and (42) approach zero as
  $r_-\rightarrow r_+$ for $\tilde{d} > 2$ and $\tilde{d}=1$.
 Hence all these second moments diverge in the extremal limit 
 ($r_-= r_+$). For $\tilde{d}=2$, all these second moments
are finite even in the extremal limit. The divergence of second
moments for $\tilde{d}\ne 2$ implies that the extremal limit is on
 a critical point. In addition, it should be pointed out that,
 in this ensemble, all second moments
are finite at the points of Davies.

(ii) {\it Canonical ensemble}. In this case, the black
 $p$-branes can only
exchange the heat with  surroundings. The variation of 
 proper Massieu function is
\begin{equation}
d\Psi _2=dS-d(\beta _H M)=-Md\beta _H-\varphi dQ.
\end{equation}
Here the intrinsic variables become $x_i=\{\beta _H, Q\}$, and the
conjugate
 variables $X_i=\{-M, -\varphi\}$. The eigenvalues are
\begin{eqnarray}
\lambda _{2\beta_H}&=&-\left(\frac{\partial \beta _H}{\partial M}
\right)_Q
=\frac{\beta ^2_H}{C_Q},\\
\lambda _{2q}&=&-\left(\frac{\partial Q}{\partial \varphi}
\right)_{\beta _H}=-\frac{K_{\beta _H}}{\beta _H},
\end{eqnarray}
where
\begin{equation}
K_{\beta_H}\equiv\beta_H\left(\frac{\partial Q}{\partial
 \varphi}\right)_{\beta _H}
=\frac{\tilde{d}\Omega _{\tilde{d}+1}(r_+r_-)^{\tilde{d}/2}}{\pi}
\frac{[1-
(\tilde{d}-1)(r_-/r_+)^{\tilde{d}}]}{[1-(r_-/r_+)^{\tilde{d}}]}.
\end{equation}
From (34), we obtain  second moments as follows:
\begin{eqnarray}
\langle\delta M \delta M\rangle&=&k_BT^2C_Q,\nonumber\\
\langle\delta S \delta M\rangle&=&k_BTC_Q,\nonumber\\
\langle\delta \varphi \delta \varphi\rangle&=&-k_B
\frac{\beta _H}{K_{\beta _H}},
\nonumber\\
\langle\delta \Phi \delta \Phi\rangle&=&-k_B\frac{T}
{K_{\beta _H}},\nonumber\\
\langle\delta S \delta S\rangle&=&k_BC_Q.
\end{eqnarray}
For $\tilde{d} > 2$ and $\tilde{d}=1$, the eigenvalues
 $\lambda _{2\beta_H}$
and $\lambda _{2q}$ diverges in the extremal limit. 
Thus all second moments are finite,  and approach zero in the
 extremal limit.  For $\tilde{d}=2$, all these second
 moments are finite even if the extremal limit is taken.
We wish to comment that contrary to the microcanonical ensemble, 
all second moments diverge at the  Davies' points in canonical
 ensemble.

(iii) {\it Grand canonical ensemble}. In this case the black 
$p$-branes can  exchange the heat with surroundings. Also they can 
do work on the surroundings.
 The variation of the proper Massieu function  is
\begin{equation}
d\Psi _3=d\Psi _2+d(\varphi Q)=-Md\beta _H +Qd\varphi.
\end{equation}
Here we read the intrinsic variables $x_i=\{\beta _H, \varphi \}$, 
and the conjugate variables $X_i=\{-M, Q\}$. The corresponding 
eigenvalues lead to
\begin{eqnarray}
\lambda _{3\beta _H}&=&-\left(\frac{\partial \beta _H}{\partial M}
\right)_{\varphi}=\frac{\beta ^2_H}{C_{\varphi}},\\
\lambda _{3\varphi}&=&\left(\frac{\partial \varphi}{\partial Q}
\right)_{\beta _H}=\frac{\beta _H}{K_{\beta _H}},
\end{eqnarray}
where
\begin{eqnarray}
C_{\varphi}&\equiv &\left(\frac{\partial M}{\partial T}
\right)_{\varphi}
  =\frac{\Omega _{\tilde{d}+1}r_+^2}{4}\left[1-\left(
  \frac{r_-}{r_+}
\right)^{\tilde{d}}
 \right]^{(\tilde{d}+2)/2\tilde{d}}[(\tilde{d}+1)r_+^{\tilde{d}-1}-
r_-^{\tilde{d}-1}{\cal G}]  \nonumber\\
&\times&\left \{\left [-1+\frac{\tilde{d}}{2}\left (\frac{r_-}{r_+}
\right)^{\tilde{d}}\right ]
-\frac{\tilde{d}-2}{2}\left(\frac{r_-}{r_+}\right)^{\tilde{d}-1}
{\cal G}\right \}^{-1},
\end{eqnarray}
and
$$ {\cal G}=\frac{\tilde{d}-2}{2}\left(\frac{r_-}{r_+}\right)\left[
\frac{\tilde{d}}{2}-
\left(\frac{r_-}{r_+}\right)^{\tilde{d}}\right]^{-1}.   $$
In this ensemble, the nonvanishing second moments are
\begin{eqnarray}
\langle\delta M \delta M\rangle&=&k_BT^2C_{\varphi}, \nonumber\\
\langle\delta Q \delta Q\rangle&=&k_BTK_{\beta _H},\nonumber\\
\langle\delta S \delta S\rangle&=&k_B(C_{\varphi}+\varphi ^2
 TK_{\beta _H}),\nonumber\\
\langle\delta S \delta M\rangle&=&k_BTC_{\varphi}, \nonumber\\
\langle\delta Q \delta S\rangle&=&-k_B\varphi ^2 T K_{\beta _H}.
\end{eqnarray}
For $\tilde{d} > 2$, we have $\lambda _{3\beta _H}\rightarrow
 \infty$
 and $\lambda _{3\varphi}\rightarrow 0$ in the extremal limit.
 Thus, $\langle\delta M \delta M\rangle$ and $\langle\delta S 
 \delta M\rangle$ are finite, while
 $\langle\delta Q \delta Q\rangle$, $\langle\delta Q \delta S
 \rangle$,
and $\langle\delta S \delta S\rangle$ diverge under the extrtemal
 limit. For $\tilde{d}=1$,
 $\langle\delta S \delta S\rangle$ and $\langle\delta Q
 \delta S\rangle$
 diverge but others are finite in the extremal limit. 
For $\tilde{d}=2$,
all these second moments are finite  again. In this ensemble,
all second
 moments are also finite at the points of Davies.

\subsection{Scaling laws and critical exponents}

Up to now we have calculated some second moments for
different fluctuation modes. It is shown that the thermodynamic
 fluctuations and  second moments of black
 $p$-branes are resulted in different, according to the different 
 environments. In order
to discuss the critical behavior of the isolated black $p$-branes
 as in [18,19], it is appropriate to choose the microcanonical
ensemble. Here the proper Massieu function is given by the 
entropy of black $p$-branes.
 From (44) it follows that all second moments diverge in the 
 extremal limit.  Therefore it is suggested that the extremal 
  black $p$-branes has a critical point,
as in the cases of the Kerr-Newman black holes and BTZ black
 holes. According to the ordinary thermodynamics, it is 
 conjectured that
the extremal and nonextremal black $p$-branes are two 
different phases.  As is well known, the extremal black configurations
 are very  different from the nonextremal black configurations
 in many respects. For example, there exists the difference of the 
topological structures between the extremal  and nonextremal black 
holes [32,33]. In the Euclidean manifolds of black holes, a conical
 singularity exists at the event horizon for a nonextremal black hole
 and results in the periodicity of the Euclidean time in order to 
remove the singularity, which gives us the inverse Hawking 
temperature of the black hole. The periodicity is absent for 
extremal black holes. This gives rise to the argument of Hawking
 {\it et al} [32] that an extremal black hole can be in the  
equilibrium with an arbitary temperature heat bath.  A nonextremal
 black hole can be described by the black hole thermodynamics. But 
the thermal description fails for some extremal black holes [15].
 In addition, the mechanism of radiations is also different for the
 extremal  and nonextremal black holes. There exist two
 kinds of radiations,  the Hawking
evaporation and the superradiation, in the nonextremal black holes. 
For an extremal black hole only the superradiation is left because 
the Hawking temperature is zero (although the Hawking temperature of
some dilaton black holes
 becomes infinity under the extremal limit,
the Hawking evaporation is still killed by an infinit gap outside the 
black holes [15]). In particular, it has been shown that some extremal 
black configurations are supersymmetric and the supersymmetry is absent
 for their nonextremal counterparts. Some shown examples are
the Reissner-Nordstr\"om black holes [34], BTZ black holes [35], 
dilaton black holes [36], and black strings and $p$-branes [21, 22, 37]. 
Therefore, we argue that the extremal limit is a critical point and a 
  second-order 
phase trasition takes place from the extremal to nonextremal black 
$p$-branes. The
 extremal and nonextremal black $p$-branes are 
in the different phases. The extremal black $p$-branes are in the 
disordered
phase and the nonextremal in the ordered phase because the extremal 
black $p$-branes have higher symmetry than the nonextremal. 
Accompanying the phase transition is the breaking of the
 supersymmetry.

 In the phase transition theory of ordinary
thermodynamics, the order parameter is a very important physical 
quantity. In general, the order parameter is chosen among  the 
extensive 
variables or their densities [38, 39]. One of the extensive variables, 
for example, $a_j=X$ is related to the order parameter $\eta=X-X_1$, 
where $X_1$ is the equilibrium value of $X$ in the disordered phase. 
Thus, the order parameter is equal to zero and is defferent from zero
 in the ordered phase. In the ordinary thermodynamic system, 
the extensive 
variables are different and the intensive variables are 
common in the two phases. But the intensive quantities are 
different and the extensive ones are common in the two
 phases of black holes. 
For the nonextremal $p$-branes in microcanonical ensemble,
 the intensive quantities $(\beta_H,\varphi)$ are different,
 while the extensive ones $(M,Q)$ are same at the $r_+$
  and $r_-$. We  can thus regard  
 the differences
$\eta _{\beta _H}=\beta _{H+}-\beta _{H-}$ and
 $\eta _{\varphi}=\varphi _+-\varphi _-$ as the order parameters
of the black $p$-branes. Here the suffixes ``$+$'' and ``$-$'' mean
 that the quantity is taken at the $r_+$ and
 $r_-$, respectively. When the extremal limit is approached, one can
  observe the critical bebavior of order
parameters. Differing from the order parameters 
in the usual thermodynamic systems, the 
order parameters of the $\tilde{d} >2$ black $p$-branes diverge 
under the extremal
 limit. This is because the
critical temperature is zero in this phase transition. On the other hand, 
the second-order derivatives of 
entropy with respect
to the intrinsic variables are given by the inverse eigenvalues
\begin{eqnarray}
\bar{\zeta}_1&\equiv &\left (\frac{\partial ^2S}{\partial M^2}\right)_Q=
\lambda _{1m}^{-1}=-\frac{\beta ^2_H}{C_Q},\\
\bar{\zeta}_2&\equiv&\left (\frac{\partial ^2 S}{\partial Q^2}\right)_M=
\lambda _{1q}^{-1}=-\frac{\beta _H}{K_M}.
\end{eqnarray}
According to  Kaburaki [18], the critical exponents of
 these
 quantities can be given as follows,
\begin{eqnarray}
\bar{\zeta}_1&\sim&\varepsilon _M^{-\alpha}\ \ \ ({\rm for\ Q\ fixed}),
            \nonumber\\
         &\sim&\varepsilon _Q^{-\psi}\ \ \ ({\rm for\ M\ fixed}),\\
\bar{\zeta}_2 & \sim & \varepsilon ^{-\gamma}_M \ \ \
             ({\rm for\ Q\ fixed }),      \nonumber \\
          & \sim &\varepsilon ^{-\sigma}_Q \ \ \ ({\rm for\ M\ fixed}),
\\
\eta _{\varphi}&\sim &\varepsilon ^{\beta}_M \ \ \ ({\rm for\ Q\ fixed}),
                  \nonumber\\
      &\sim &\varepsilon ^{\delta ^{-1}}_Q \ \ \ ({\rm for\ M\ fixed}),
\end{eqnarray}
where $\varepsilon _M$ and $\varepsilon _Q$ represent the infinitesimal
deviations of $M$ and $Q$ from their limit values. Making use of Eqs.(55)
and (56), we obtain
\begin{equation}
\alpha=\psi=\gamma=\sigma=(3\tilde{d}-2)/2\tilde{d},\ \ \beta=
\delta ^{-1}=-(\tilde{d}-2)/2\tilde{d}.
\end{equation}
The critical exponents $\beta$ and $\delta ^{-1}$ are negative for
 $\tilde{d} > 2$, which shows  that in this case the order 
 parameter
$\eta_{\varphi}$
 diverges in the extremal limit ($r_- \to r_+$). We note that these  
 exponents are different from those of
Kerr-Newman black holes and BTZ black holes [19]. But they still satisfy 
the scaling laws of the ``first kind'' as
\begin{eqnarray}
&&\alpha +2\beta +\gamma =2, \nonumber \\
&&\beta(\delta-1)=\gamma, \nonumber \\
&&\psi (\beta +\gamma)=\alpha.
\end{eqnarray}
That the scaling laws (61) hold in the black $p$-branes is related to 
the fact  that the  black $(d-1)$-brane
entropy is a homogeneous function, satisfying
\begin{equation}
S(\lambda M, \lambda Q)=\lambda ^{(\tilde{d}+1)/\tilde{d} }S(M,Q),
\end{equation}
where $\lambda$ a positive 
constant. This can be seen clearly from Eqs. (21), (29) and (30).

In the ordinary thermodynamic system, the correlation function is an
important quantity to extract the information for  phase transitions. 
Near the critical points,
this function takes generally the form  [38,39]
\begin{equation}
G(r)\sim \frac{\exp (-r/\xi)}{r^{\bar{d}-2+\eta}},
\end{equation}
where $\eta$ is the Fisher's exponent, $\bar{d}$ is the effective
spatial
dimension of the system under consideration, and $\xi$ is the
correlation
 length. Applying this to our system, we have
\begin{eqnarray}
\xi &\sim & \varepsilon ^{-\nu}_M \ \ \ ({\rm for\ Q\ fixed}),
\nonumber\\
    &\sim &\varepsilon ^{-\mu}_Q \ \ \ ({\rm for\ M\ fixed}).
\end{eqnarray}
Further these critical exponents satisfy the scaling laws of the
``second kind'',
\begin{equation}
\nu (2-\eta )=\gamma, \ \ \nu \bar{d}=2-\alpha, \ \ \mu (\beta
 +\gamma)=\nu.
\end{equation}
Due to the absence of quantum theory of gravity, by now the
correlation function
of quantum black holes has yet been unclear. Instead,  we
 use the
correlation function of scalar fields
to get the quantum aspects of  black $p$-branes [19].  
Traschen [40]  has recently studied
the behavior of a massive charged scalar
  field on the background of Reissner-Nordstr\"{o}m black holes. 
 Traschen has found that the spacetime
   geometry near the horizon of the extremal Reissner-Nordstr\"{o}m 
black holes has a scaling symmetry, which is absent for the 
nonextremal holes, a scale being introduced by the surface gravity. 
The scaling symmetry results in that an 
external source has a long range influence on the extremal 
background as $y^{-1}$, compared to a correlation length scale 
which falls off exponentially fast in the 
case of nonextremal holes, like $e^{2\kappa y}$, where $\kappa$ 
is the surface gravity and $y$ is the usual tortoise coordinate. 
 Therefore the inverse surface
gravity is regarded as the correlation length. Here we assume that this
also holds
for our black $p$-branes. Making use of the surface gravity
of black $p$-branes, we obtain
\begin{equation}
\nu=\mu=(\tilde{d}-2)/2\tilde{d}.
\end{equation}
Substituting these into (65), we have
\begin{equation}
\eta =-(\tilde{d}+2)/(\tilde{d}-2),\ \ \bar{d}=(\tilde{d}+2)/
(\tilde{d}-2).
\end{equation}
The two critical exponents are also different from the BTZ
 black holes [19].
 In particular, for $\tilde{d}=1$, the effective dimension
$\bar{d}=-3$. The appearance of the negative effective
 dimensionality is due to that the surface gravity of $\tilde{d}=1$
 black $p$-branes is divergent under the extremal limit. 
 More recently, such a negative exponent scaling has also been 
found by Klebanov and Tseytlin [41] in the intersecting M-brane 
configurations. They argued that this scaling can be explained 
by the dynamical $p$-branes. In addition, unlike in the BTZ black 
holes, in which the effective dimension $\bar{d}=1$, from (67) we
 see that generally the effective dimension of these dilaton
 $p$-branes is not an integer.

So far, we have investigated the critical behavior of black
$(d-1)$-branes with ``electric'' charge. The extremal limit of
$(d-1)$-branes has a critical point and a second-order phase 
transition
may take place from the extremal to nonextremal black $p$-branes.
  It is already shown that these extremal black 
$p$-branes are  supersymmetric and the supersymmetry is always
 absent for their nonextremal counterparts [21, 22, 37]. 
Therefore, that a phase transition occurs from extremal to
 nonextremal black $p$-branes agrees with the fact of symmetric 
changes.  We have argued that the extremal black 
$p$-branes and nonextremal black $p$-branes are two different
 phases, and the extremal
 black $p$-branes are in the disordered phase and 
the nonextremal black $p$-branes in the ordered phase.
However $\tilde{d}=2$ is a special case. In this case, the extremal
limit does not  have a critical point because the thermodynamic
 quantities and fluctuations have good bahavior even in the
extremal limit. This situation is similar to that of the $a=1$ 
charged dilaton black holes. However, if one considers other fields
and/or quantum corrections in the theory, we argue that the extremal 
limit will have a critical point.  This will be seen below.

\section{Charged dilaton black holes}

It is well known that the behavior
 of black holes is drastically changed because of the presence of  
 dilaton fields. In order to understand 
the effect of the dilaton field on the critical behavior of 
black holes, we choose the four dimensional  dilaton action [42],
\begin{equation}
S=\frac{1}{16\pi}\int d^4x\sqrt{-g}[R-2(\nabla \phi)^2
-e^{-2a\phi}F^2],
\end{equation}
where  $F$ is the
Maxwell field. The coupling constant $a$ governs the interaction
between the dilaton field and the Maxwell field. For $a=1$, 
the action corresponds to the
low-energy approximation of superstring theory. The charged 
 black hole solutions  are
\begin{eqnarray}
&&F=\frac{e^{2a\phi}}{R^2}Q, \nonumber\\
&&e^{2a\phi}=\left(1-\frac{r_-}{r_+}\right)^{2a^2/(1+a^2)},
 \nonumber\\
&&ds^2=-A^2(r)dt^2+ A^{-2}(r)dr^2+R^2(r)d\Omega ^2,
\end{eqnarray}
where
\begin{eqnarray}
&&A^2(r)=\left(1-\frac{r_+}{r}\right)\left(1-\frac{r_-}{r}
\right)^{(1-a^2)/(1+a^2)},\nonumber\\
&&R^2(r)=r^2\left(1-\frac{r_-}{r}\right)^{2a^2/(1+a^2)}.
\end{eqnarray}
Two parameters $r_+$ and $r_-$ are related to 
 the  mass $M$ and electric charge $Q$,
\begin{equation}
 2M=r_++\frac{1-a^2}{1+a^2}r_-, \ \ Q^2=\frac{r_-r_+}{1+a^2}.
\end{equation}
The Hawking temperature $T$ and entropy $S$ of the dilaton 
black holes are
\begin{eqnarray}
&&T=\frac{1}{4\pi r_+}\left(1-\frac{r_-}{r_+}\right)
^{(1-a^2)/(1+a^2)}, \\
&&S=\pi r_+^2\left(1-\frac{r_-}{r_+}\right)^{2a^2/(1+a^2)}.
\end{eqnarray}
The heat capacity of the dilaton black holes ($C_Q=
(\partial M/\partial T)_Q)$ is found to be
\begin{equation}
C_Q=-\frac{2\pi r_+^2\left[1-\frac{1-a^2}{1+a^2}\frac{r_-}
{r_+}\right]\left[1-\frac{r_-}
{r_+}\right]^{2a^2/(1+a^2)}}{\left[1-\frac{3-a^2}{1+a^2}
\frac{r_-}{r_+}\right]}.
\end{equation}
From Eqs.(72)-(74) we easily find that the thermodynamic
 quantities depend explicitly on the coupling constant $a$. In the
 extremal limit ($r_-\rightarrow r_+$), the Hawking temperature is
zero, finite, and divergent when $a < 1$, $a=1$, and 
$a > 1$, respectively. But
the entropy  always vanishes.  For $a > 1$, the heat capacity
is always  negative and approaches  zero in the extremal limit. 
When $a=1$, the heat capacity
$C_Q=-8\pi M^2$, independent of the electric charge $Q$. 
Furthermore, for $a < 1$, it is possible to have
\begin{equation}
1-\frac{3-a^2}{1+a^2}\frac{r_-}{r_+}=0,
\end{equation}
which means that $C_Q$ diverges in this case. These points 
correspond to the turning points of Davies. At these points, the
heat
capacity encounters with an infinite discontinuity.
With  (71)-(73), we have the first law of
thermodynamics
\begin{equation}
dM=\beta _H^{-1}dS+\Phi dQ,
\end{equation}
where $\beta ^{-1}_H=T $ and $\Phi=Q/r_+$ is the electric potential
at the horizon $r_+$. The mass formula of the holes is
\begin{equation}
M=2\frac{\kappa}{2\pi}\frac{\Sigma}{4}+\Phi Q,
\end{equation}
where $\kappa$ ($\Sigma$) are the surface gravity 
(the horizon area) of the black holes.

Applying the results for  the $p$-branes in the microcanonical
ensemble to this system, the eigenvalues and nonvanishing second
 moments  are still given by Eqs. (41), (42) and (44),
respectively. But the heat capacity $C_Q$  should be replaced by the
 expression (74), and $K_M$ is 
\begin{eqnarray}
K_M=\left[\frac{1+a^2}{1-a^2}-\frac{r_-}{r_+}\right] \left [\frac{1}
{r_+}\left(\frac{1+a^2}{1-a^2}+\frac{r_-}{r_+}\right)
  + \frac{4r_-^2}{(1+a^2)r_+^3}\left(1-\frac{r_-}{r_+}\right)^{-1}
\right]^{-1}.
\end{eqnarray}
Again we find that  two eigenvalues of $\beta _H$
and $\varphi$ approach zero in the extremal limit, and hence all
 these second moments (44) diverge.  For the case of $a=1$, the 
 eigenvalues and
second moments  are always finite. Therefore the all extremal 
 black holes except $a=1$ have   critical points. Similarly, 
 we can obtain  relevant critical exponents
\begin{equation}
\alpha =\psi=\gamma=\sigma=\frac{2}{1+a^2}, \ \ \beta =\delta ^{-1}=
-\frac{1-a^2}{1+a^2}.
\end{equation}
Naturally, these  also satisfy the scaling laws of
the ``first kind'' (61), because the entropy (73) is a homogeneous 
function, which satisfies
\begin{equation}
S(\lambda M, \lambda Q)=\lambda ^2 S(M, Q)
\end{equation}
with a positive constant $\lambda$. From the surface gravity and
 the scaling laws of the ``second kind'', we have
\begin{equation}
\nu=\mu=\frac{1-a^2}{1+a^2},
\end{equation}
and
\begin{equation}
\eta= -\frac{2a^2}{1-a^2}, \ \ \bar{d}=\frac{2a^2}{1-a^2}.
\end{equation}
Here the negative effective dimension $\bar{d}$ appears again 
when $a > 1$.
 This arises from  the same reason as in the
 $(d-1)$-branes
 with $\tilde{d}=1$, because  the surface gravity diverges in
 the extremal limit.
 For $a=0$, the charged dilaton black holes reduce
 to the Reissner-Nordstr\"{o}m black holes. After some calculations,
we have
\begin{eqnarray}
&&\alpha =\psi=\gamma=\sigma=3/2, \ \ \beta =\delta ^{-1}=-1/2,
 \nonumber\\
&&\nu=\mu=1/2, \ \ \eta=-1, \ \ \bar{d}=1.
\end{eqnarray}
These  exponents are exactly the  same as those of the BTZ black
holes [19].
The critical behavior of the extremal
 dilaton black holes in the microcanonical ensemble is in complete
agreement
with the result of nonequilibrium thermodynamic fluctuations [14].
The thermodynamic fluctuations of dilaton black holes in the
canonical
 ensemble and grand canonical ensemble can also be given 
 similarly as in the black $p$-branes. In the canonical ensemble, 
 the divergent point of second moments is
 the point of Davies (75).

From the above, we find that the extremal  black $p$-branes
and dilaton black holes have  critical points, and the 
relevant critical exponents satisfy the scaling laws. This 
is the important aspect of phase transitions.
On the other hand, the extremal $\tilde{d}=2$-branes 
and extremal $a = 1$ dilaton black holes
 have not the critical points and show no features of phase 
 transitions. In addition,
the $\tilde{d}=1$ black branes and $a>1$ dilaton black holes 
have the negative effective spatial dimension.
Concerning this problem, although Klebanov 
and Tseytlin [41] have 
made some discussions in the intersecting M-brane configurations, 
to understand completely the negative scaling in all appearing 
cases, further investigation is explicitly needed.
 For example, we need 
to investigate  further whether or not these extremal black
configurations have  the
scaling symmetry and the external source has the long range influence
 on these extremal black configuration backgrounds. 
 For the $\tilde{d}=2$ black branes and $a=1$
dilaton black holes,  we would like to point out that if one further
considers other fields in the string theories, the extremal limit
 should be a critical point, because the actions (1) and (68) are 
the simpler approximations 
of the lower-energy actions of string theories. Recall that the 
action (68) with parameter $a=1$ is just the lower-energy limit 
of the heterotic superstring
 theory,  in order to study the relationship between extremal
 black configurations and strings, it is necessary to consider the 
dilaton black holes with other fields and/or quantum corrections. 
In the next section, we will study the $a=1$ charged 
dilaton black hole with a moduli field.

\section{ Charged dilaton black holes with moduli fields}

The actions (1) and 
(68) are the first approximations of lower-energy actions of
 string theories. For the large
 mass black holes ($M >> M_{\rm pl}$), these should be a good 
approximation. But, here we are considering the quantum effects 
of these configurations, in
 order to understand the critical behavior of these black
 configurations in the string theories, the effects of other
 fields  in the string theories and/or 
quantum corrections to the lower-energy action of string theories
 should be investigated.
 In this section,
 we study the charged dilaton black holes with moduli fields. 
 These come
from the low-energy action of heterotic theories by compactification
from ten to four dimensions. Here we include one single
 modulus which describes the radius of a compactified space.
The action is
\begin{equation}
S =\frac{1}{16\pi} \int d^4 x\sqrt{-g}\left [R-2(\nabla \phi)^2-
   \frac{2}{3}(\nabla \sigma )^2  \right.
  \left.-e^{-2\phi}F^2-e^{-2q\sigma/3}F^2 \right ].
\end{equation}
This is a modification of the action (68) with $a=1$ by including a 
moduli field
$\sigma$ with a coupling constant $q$. The black hole solutions in
 the action (84) are found already by Cadoni and Mignemi [43]. The
metric takes the form in Eq. (69) but with the substitutions 
\begin{eqnarray}
&& A^2(r)=\left[1-\frac{r_+}{r}\right]\left[1-\frac{r_-}{r}
\right]^{3/(2q^2+3)}, \nonumber\\
&& R^2(r)=r^2\left[1-\frac{r_-}{r}\right]^{2q^2/(2q^2+3)}, 
\nonumber\\
&&e^{-2\phi}=\left[1-\frac{r_-}{r}\right]^{2q^2/(2q^2+3)}, 
\nonumber\\
&&F=Q\varepsilon _2, \ \ \sigma =3\phi /q.
\end{eqnarray}
This is a two-parameter black hole solution with magnetic charge
 $Q$. The electrically charged solutions can be obtained by the 
 duality transformation.
The ADM mass $M$ and charge $Q$ have the relations with two 
parameters $r_-$ and $r_+$,
\begin{eqnarray}
&& 2M=r_++\frac{3}{2q^2+3}r_-,\\
&& Q^2=\frac{q^2 +3}{2q^2+3}r_-r_+.
\end{eqnarray}
The Hawking temperature and entropy of the black holes are easily 
calculated as
\begin{eqnarray}
&&T=\frac{1}{4\pi r_+}\left[1-\frac{r_-}{r_+}\right]^{3/(2q^2+3)},
 \\
&&S=\pi r_+^2\left[1-\frac{r_-}{r_+}\right]^{2q^2/(2q^2+3)}.
\end{eqnarray}
The heat capacity is
\begin{equation}
C_Q=-2\pi r_+^2\left[1-\frac{r_-}{r_+}\right]^{2q^2/(2q^2+3)}
    \left[1-\frac{3}{2q^2+3}\frac{r_-}{r_+}\right]
    \left[1-\frac{2q^2+9}{2q^2+3}\frac{r_-}{r_+}\right]^{-1}.
\end{equation}
From Eqs. (88)-(90), it is clear that the thermodynamics of 
the $a=1$ charged dilaton black holes is changed drastically 
due to the presence of the moduli field. In this case,  
the Hawking temperature and entropy
 are zero in the extremal limit ($r_- \to r_+$). This means that 
 the extremal limit corresponds to a stable nondegenerate ground 
 state. In addition, the
entropy (89) is still a homogeneous function of the form (80). 
When one goes to the extremal limit, the heat capacity (90) 
approaches  zero. But this
  diverges and changes its sign,  when goes to points of Davies 
  which   satisfy the relation
\begin{equation}
1-\frac{2q^2+9}{2q^2+3}\frac{r_-}{r_+}=0.
\end{equation}
 Now we are in a position to consider the
 equilibrium thermodynamic fluctuations of this model in 
 the microcanonical ensemble. These eigenvalues and second 
moments are still given by Eqs.
 (41), (42) and (44), respectively. In this case, $C_Q$ is given 
by Eq. (90) and $K_M$ is
\begin{equation}
K_M =\left[\frac{2q^2+3}{3}-\frac{r_-}{r_+}\right]
\left\{\frac{1}{r_+}\left[\frac{2q^2+3}{3}+\frac{r_-}{r_+}\right]
+\frac{2q^2+6}{2q^2+3}\frac{r_-^2}{r_+^3}
\left[1-\frac{r_-}{r_+}\right]^{-1}\right\}^{-1}.
\end{equation}
 The eigenvalues of
 the fluctuation modes become zero and all second moments 
 diverge at the extremal
limit. Similarly, corresponding critical  exponents are easily 
obtained and read as
\begin{equation}
\alpha =\psi=\gamma=\sigma=\frac{2q^2+6}{2q^2+3}, \ \ \beta =
\delta ^{-1}=-\frac{3}{2q^2+3}.
\end{equation}
The critical exponents of the correlation length are
\begin{equation}
\nu=\mu=\frac{3}{2q^2+3},
\end{equation}
and from the scaling laws of the ``second kind'', we have
\begin{equation}
\eta=-2q^2/3, \ \ \bar{d}=2q^2/3.
\end{equation}
Thus this model  has a critical point because of the presence
of the moduli fields. We argue that the extremal black
configuration in the superstring theories has the critical point, 
and related
exponents satisfy the scaling laws. In the canonical ensemble, all
second moments are finite except the points of Davies (91).
It is worth noting that when $q=0$, the solution (85) does not
 reduce to the $a=1$
dilaton black  holes, but the Reissner-Nordstr\"{o}m black hole.
 In the string frame ($ds^2_S=e^{2\phi}ds^2$), the metric  (85)
becomes
\begin{equation}
ds^2_S=-\left [1-\frac{r_+}{r}\right]^{6/(2q^2+3)}dt^2
  +\left[1-\frac{r_+}{r}\right]^{-2}dr^2 +r^2d\Omega ^2
\end{equation}
in the extremal limit.
The geometry of the $t=$constant  surface  is exactly
identical to that of a static slice in the extremal 
Reissner-Nordstr\"{o}m spacetime.
 Here is a semi-infinite throat attached to an asymptotically flat
  region. This supports that the extremal  dilaton black holes
with moduli field has a critical point, as in the case of the
Reissner-Nordstr\"{o}m black holes.

\section{Conclusion and discussion}

In this work we investigated  the critical
 behavior for
the  black $p$-branes and four dimensional  dilaton
 black holes with (and without) a moduli field. We 
confirmed that the thermodynamic fluctuations depend on 
the kind  of ensembles. Hence three  ensembles for a 
self-gravitating system are not equivalent.
 In the microcanonical ensemble, all  second moments of
  extremal black configurations diverge. Some second moments 
  diverge at the points of
 Davies in the canonical ensemble. But the points of Davies are
  actually the turning points, which  reveal to the changes of 
  stability and have nothing to do with phase transition.
The extremal  black configurations has a critical point.
 And the
phase transition takes place from the extremal to nonextremal black
configurations. The related critical exponents satisfying the scaling
 laws are obtained. Note that the entropy of these black 
configurations  takes a property of
 the  homogeneous function.

A phase transition is always accompanied by the symmetry
changes in the ordinary thermodynamic systems.
 As is well known, some  extremal black
configurations are  supersymmetric and the supersymmetry is absent for
 nonextremal ones. Thus it is suggested that the phase transition  from
the extremal to nonextremal black
configurations takes the change of supersymmetry.
We argued that the extremal and nonextremal black 
configurations are two different phases. The extremal black 
configurations are in the disordered phase and the nonextremal 
in the ordered phase because the extremal black configurations have
 higher symmetry than the nonextremal ones.
 For the 
$\tilde{d}=1$ black $p$-branes and $a >1$ charged dilaton 
black holes, the 
effective spatial dimension is negative. This is because of the 
divergence of surface gravity for these extremal black 
configurations. In order to understand the origin of the negative 
scaling, further study beyond the scope of this paper should be
 required. The recent work of Klebanov and Tseytlin [41] seems to
  point out a promising direction. In addition,  to further 
  investigate the critical behavior of these special extremal 
  configurations, it is of interest to study that
the behavior of quantum fields on the these extremal 
configuration backgrounds, as was done by Traschen [40] on 
the Reissner-Nordstr\"{o}m background, to see 
whether or not the geometry of these extremal configurations 
has the scaling symmetry, and whether the external sources have 
the long range influence on these extremal backgrounds.
The extremal $\tilde{d}=2$ black branes and $a=1$ dilaton black holes
do not have a critical point. 
It is, however, found a critical point for the extremal limit 
by adding a moduli field to  the $a=1$
dilaton black hole. Finally, we notice that 
the microscopic understanding of entropy for the black
holes and black $p$-branes need the condition of the
 contant dilaton field [23]. And we wish to recall  that the 
Reissner-Nordstr\"om and BTZ
black holes can be considered as the exact solutions in the string
 theories with constant dilaton
fields [42,43,44]. In this sense it is very important to investigate the 
critical behavior of  black configurations with constant 
dilaton fields \footnote{After completing this
  work, we studied further the scaling laws and critical exponents 
  of the non-dilatonic black $p$-branes [37]. It was found that the
   effective spatial dimension is one for non-dilatonic black holes 
   and black strings, and is just $p$ for the non-dilatonic
    $p$-branes [45]. This result can explain why the Bekenstein-Hawking 
    entropy may be given a
  simple world volume interpretation only for 
  the non-dilatonic $p$-branes.}.

\begin{flushleft}
{\bf \large Acknowledgments}
\end{flushleft}
R. G. Cai wishes to thank O. Kaburaki for sending some recent 
works to him, Dr. Y. K. Lau for useful discussions, 
and Professor J. D. Bekenstein for helpful comments on the 
cirtical behavior in BTZ black holes [19]. We would like to 
thank the referee for many helpful suggestions which 
improved the original version of this paper. The research of 
R.G.C. was supported in part by 
China Postdoctoral Science Foundation.

\end{document}